\documentclass[12pt]{article}
\usepackage{geometry}
\usepackage{authblk}
\usepackage{amssymb,stmaryrd}
\usepackage{amsmath}
\usepackage{amsthm}
\usepackage{amsfonts}
\usepackage{amstext}
\usepackage{algorithmic}
\usepackage{xcolor}
\usepackage{algorithm}
\usepackage[labelfont=bf]{caption}
\usepackage{caption}
\usepackage{subcaption}
\usepackage{graphicx}
\usepackage{hyperref}
\theoremstyle{definition}

\title{Cooperative metabolic resource allocation in spatially-structured systems}
\author{David S. Tourigny\thanks{dst2156@cumc.columbia.edu}}
\affil{Columbia University Irving Medical Center\\630 West 168th Street, New York, NY 10032 USA}

\begin{document}
\maketitle

\begin{abstract}
Natural selection has shaped the evolution of cells and multi-cellular organisms such that social cooperation can often be preferred over an individualistic approach to metabolic regulation. This paper extends a framework for dynamic metabolic resource allocation based on the maximum entropy principle to spatiotemporal models of metabolism with cooperation. Much like the maximum entropy principle encapsulates `bet-hedging' behaviour displayed by organisms dealing with future uncertainty in a fluctuating environment, its cooperative extension describes how individuals adapt their metabolic resource allocation strategy to further accommodate limited knowledge about the welfare of others within a community. The resulting theory explains why local regulation of metabolic cross-feeding can fulfil a community-wide metabolic objective if individuals take into consideration an ensemble measure of total population performance as the only form of global information. The latter is likely supplied by quorum sensing in microbial systems or signalling molecules such as hormones in multi-cellular eukaryotic organisms.       
     
\end{abstract}

\section{Introduction}
\label{sec:intro}

Organisms rarely exist in isolation but instead engage in dynamic interaction with their environment and peers as part of a larger population or community. Both environmental and social interactions can have a profound effect on the metabolic behaviour of an individual cell or sub-population, which must often make complex regulatory decisions while faced with uncertainty in many external factors such as nutrient availability.  Under these conditions, competition and cooperation commonly emerge due to the pressures of natural selection \cite{Axelrod81}, which further shapes the complexity of biological systems. 

Cooperation between microbial cells and populations is generally accepted to be enhanced by spatial structure in the environment, due to limited dispersal and positive assortment keeping cooperators physically clustered together with their partners \cite{Chao81,Nowak92,Hansen07,Harcombe10,Nadell10} (however, it is sometimes possible for spatial structure to disfavour cooperation, e.g., \cite{Hauert04,Oliveira14}). The same is true for metabolic partitioning between intracellular compartments of individual cells \cite{Pfeiffer01,Martin10}. Spatial structure also promotes metabolic cooperation at various levels of organisation in higher-eukaryotic organisms, prominent examples being the well-known Lactic Acid Cycle between liver and muscle \cite{Nelson05}, the Astrocyte Neuron Lactate Shuttle hypothesis in brain \cite{Belanger11}, and the proposed symbiotic production and consumption of lactate by subpopulations of cancer cells within a tumor \cite{Sonveaux08,Semenza08,Lyssiotis17}. The nature of cooperative behaviour displayed in each of these eukaryotic examples is that of `metabolic cross-feeding', which likewise forms an important form of cooperation in microbial communities \cite{Schink91,Schink02,Morris13,Song14,Harcombe14,Nadell16,DSouza18,Lin18,Smith19,Evans20}. Metabolic cross-feeding refers to the process mediated by uptake and exchange of metabolites between individuals or sub-populations, which might invest costly resources to produce metabolites that benefit others in the community rather than using them to fulfil their own metabolic requirements. Cooperative behaviours like metabolic cross-feeding are strategies that can increase the chances of survival and propagation of genes among closely related organisms. 

The benefits and exploitation of heterogenous phenotypic traits are also well-appreciated as `bet-hedging' strategies employed by cells and populations dealing with uncertainty in a fluctuating environment \cite{Seger87,Perkins09,Ackermann15,Granados17}. Under these conditions it can be considered economically sub-optimal to invest metabolic resources exclusively into a single metabolic pathway maximising the metabolic objective. Instead, it can prove advantageous to spread resource among multiple metabolic pathways in a strategic way so as to maximise the expected return, related to using the principle of maximum entropy \cite{Jaynes57,Shore80} to formulate resource allocation as an optimality problem. From an information-theoretic standpoint, the distribution of metabolic resources that best represents the current state of knowledge is the one with largest entropy \cite{Shannon48}, and therefore the maximum entropy distribution is uniquely determined as the one consistent with known constraints but expressing maximum uncertainty with respect to everything else. In biology, the maximum entropy principle has been applied to various resource allocation problems in ecology (see \cite{Harte14} for a review), stem cell multi-potency \cite{Ridden15}, and a dynamic framework for metabolic resource allocation presented in \cite{Tourigny20}. The latter was partly motivated by a recent application of the maximum entropy principle to accommodate population heterogeneity in the steady state regime of cellular metabolism \cite{DeMartino17,Fernandez19} and forms a refinement of the work by Young and Ramkrishna \cite{Young07,Young08} (see also earlier work cited therein). This framework unifies previous dynamic models for metabolism, specifically dynamic flux balance analysis (DFBA, \cite{Mahadevan02}) and related unregulated theories \cite{Provost04,Provost06}, and proves successful in describing some observed behaviours of organisms based on first principles alone, including accumulation of metabolite reserves under growth-limiting conditions. Here, the maximum entropy framework is generalised to models for cooperative metabolic resource allocation in spatially-structured systems, where exploiting heterogeneity and bet-hedging is equally relevant for organisms coping with uncertainty in a temporally-fluctuating environment \cite{Evans20}. It is shown that the cooperative extension of this principle similarly accommodates the limited capacity of an individual cell or sub-population to acquire information about the metabolic activity of other community members outside of their immediate local spatial domain.        

The remainder of this paper is organised as follows: Section \ref{sec:model} considers a spatiotemporal model for metabolism that extends the dynamic model in \cite{Tourigny20} together with a metabolic objective that governs cooperative behaviour. The resulting maximum entropy control law for cooperative resource allocation is introduced in Section \ref{sec:control}, which also explores some of its implications for metabolic cross-feeding. Finally, a concrete application to cooperative behaviour in microbial biofilms and colonies is presented in Section \ref{sec:example}. Although the content of this paper is self-contained, many of the principles build upon and complement those considered in \cite{Tourigny20},  therefore readers are encouraged to consult that previous study for reference. Additional mathematical details including the derivation of the cooperative maximum entropy control law can be found in Appendix \ref{sec:appendix}.

\section{Spatiotemporal metabolism with cooperation} 
\label{sec:model}
\begin{figure}
    \caption{Simplified metabolic reaction network model for central carbon metabolism. \label{fig:1}}
    \centering
    \begin{subfigure}[t]{\textwidth}
    \centering
        \includegraphics[width=\linewidth]{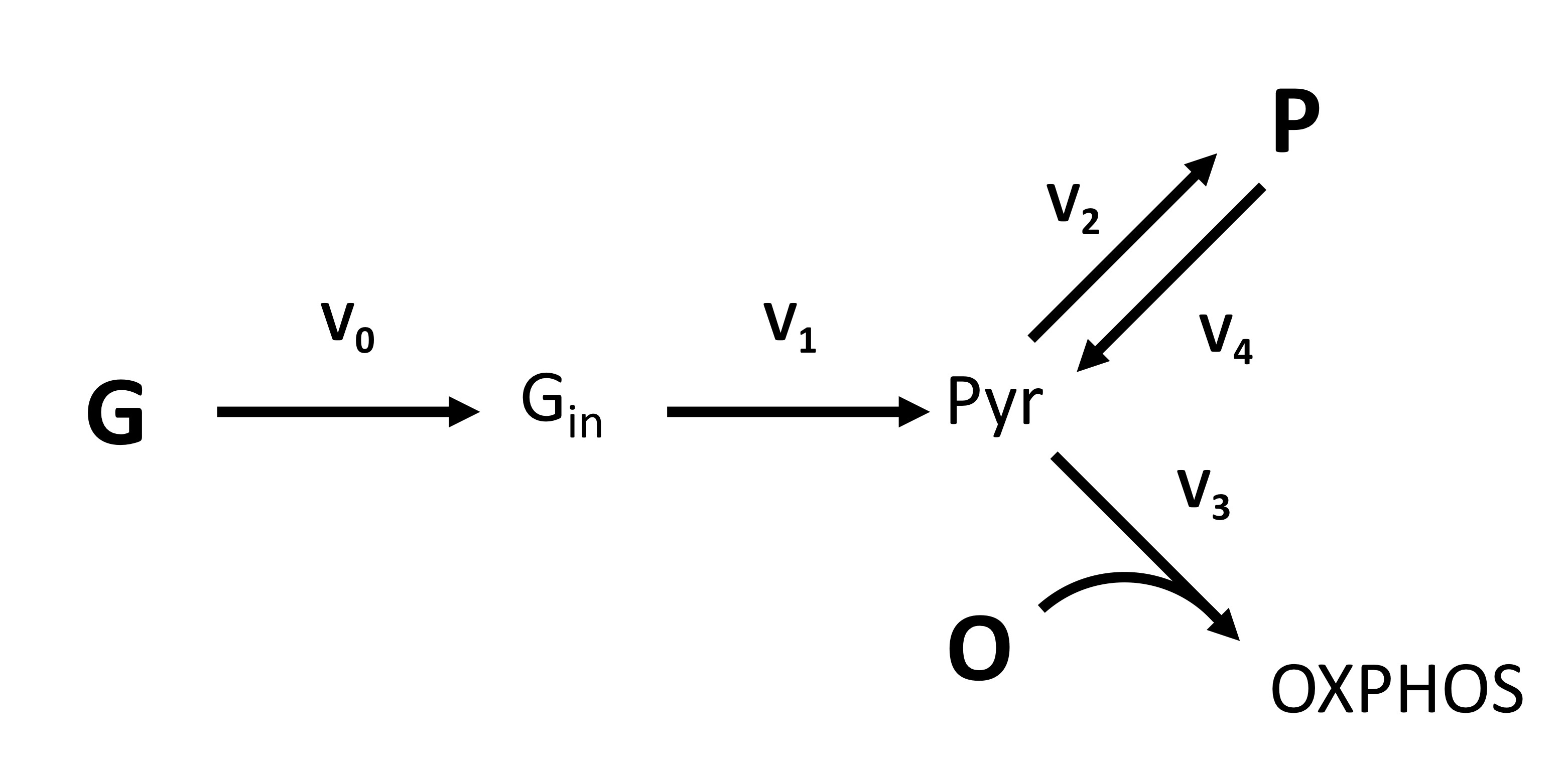}
     \caption{Diagrammatic representation of metabolic reaction network where arrowheads indicate directionality. Reactions labelled $v_0,v_2,v_3,v_4$ have unit stoichiometry while that labelled $v_1$ has stoichiometry $2$. Glucose, oxygen, and the fermentation product are slow metabolites whereas intracellular glucose and pyruvate are treated as fast metabolites. The reaction labelled $v_3$ feeds into the oxidative phosphorylation pathway, represented by OXPHOS in diagram.} \label{fig:1a} 
    \end{subfigure}
    \vspace{1cm}
    \centering
    \begin{subfigure}[t]{\textwidth}
    \centering
        \includegraphics[width=\linewidth]{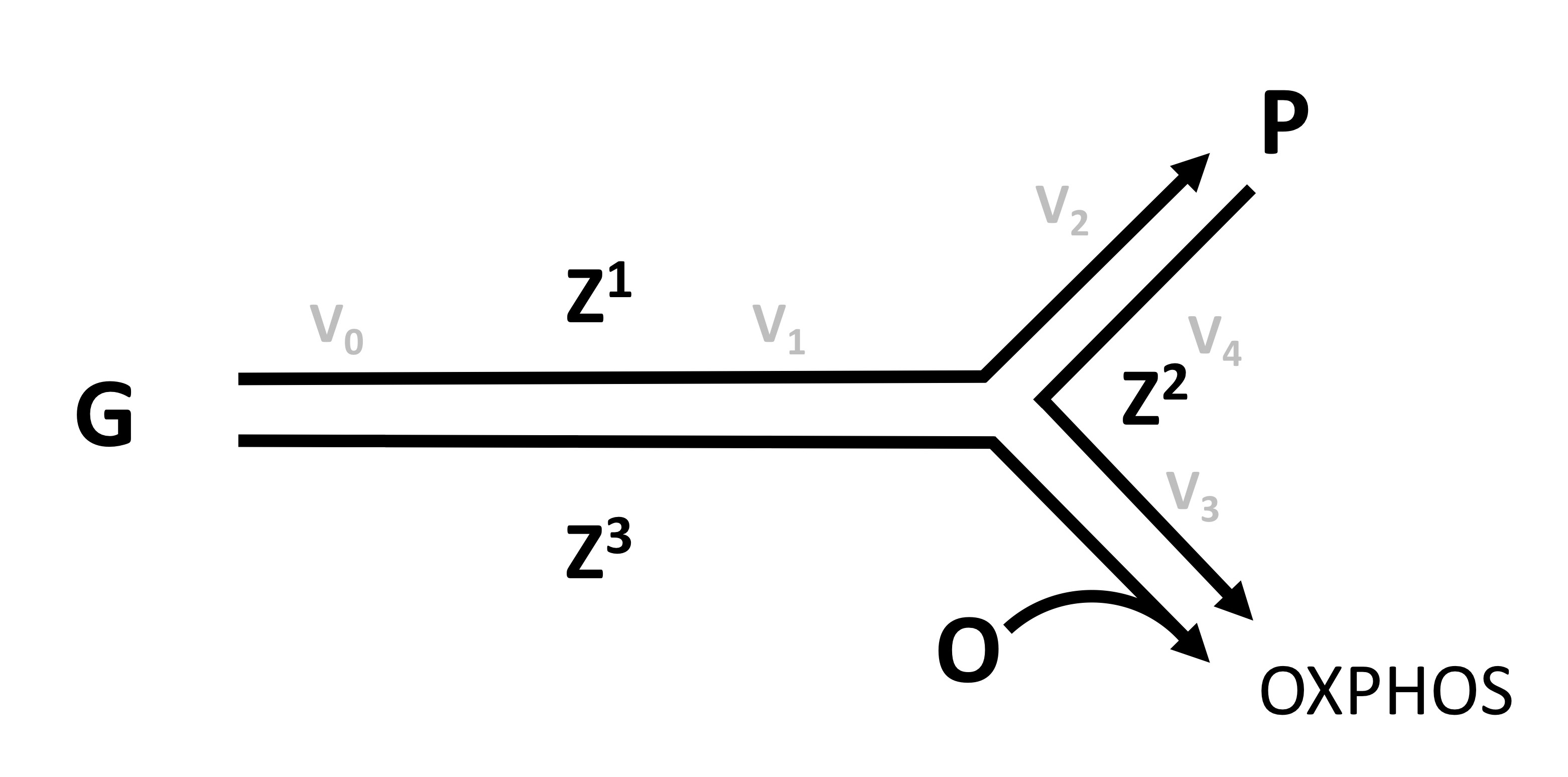} 
       \caption{Representation of three EFMs in the metabolic network by the vectors $\mathbf{Z}^1$, $\mathbf{Z}^2$, and $\mathbf{Z}^3$, corresponding to the metabolic pathways for glucose fermentation, oxidative phosphorylation of the fermentation product, and oxidative phosphorylation of glucose, respectively.} \label{fig:1b}
    \end{subfigure}
\end{figure}

In \cite{Tourigny20}, the following dynamical system was used as a model for the metabolism of a single species in batch culture
\begin{equation}
\begin{split}
\frac{d}{dt} \mathbf{m}_{ex} &= \mathbf{S}_{ex} \mathbf{v} x  \\
\frac{d}{dt} \mathbf{m}_{in} &= \mathbf{S}_{in}\mathbf{v} - \mu \mathbf{m}_{in} \\
\frac{d}{dt}x &= \mu x , \quad \mu = \mathbf{c}^T \mathbf{v} 
\end{split}
 \label{system}
\end{equation}
where $\mathbf{m}_{ex}$ ($\mbox{g}\cdot \mbox{L}^{-1}$), $\mathbf{m}_{in}$ ($\mbox{g} \cdot \mbox{L}^{-1} \cdot \mbox{gDW}^{-1} \cdot \mbox{L}$; $\mbox{gDW}$, grams dry weight) are vectors of extra- and intracellular metabolites, respectively, and $\mathbf{S}_{ex}$, $\mathbf{S}_{in}$ the corresponding portions of the stoichiometric reaction matrix $\mathbf{S}$ \cite{Varma94,Orth10,Schuster94}. The scalar variable $x$ ($\mbox{gDW} \cdot \mbox{L}^{-1}$) represents the concentration of total catalytic biomass responsible for catalysing reactions involved in its own production and interconversion of metabolites, and $\mu$ ($\mbox{h}^{-1}$) is the rate of its accumulation (i.e., growth rate) formed as the inner product of the non-negative, $N$-dimensional flux vector $\mathbf{v} = (v_1,v_2,...,v_N)^T$ ($\mbox{g} \cdot \mbox{gDW}^{-1} \cdot \mbox{h}^{-1}$) with the constant coefficient vector $\mathbf{c} = (c_1,c_2,...,c_N)^T$ ($\mbox{gDW} \cdot \mbox{g}^{-1}$). The reduction of this system to one of smaller dimension is based on the quasi-steady state assumption (QSSA) \cite{Varma94,Schuster94}, which implies metabolites can be separated into two distinct groups based on their dynamic timescales, fast and slow respectively. Under the QSSA, dynamics of {\em fast} metabolite concentrations are assumed to equilibrate rapidly over a timeframe with negligible changes in {\em slow} metabolite concentrations, which is modelled by the steady state condition, $d \mathbf{m}_{in}/dt \approx 0$, and disregarding the dilution term, $\mu \mathbf{m}_{in}$. The resulting steady state algebraic equations, $\mathbf{S}_{in} \mathbf{v} = 0$, for the $N$-dimensional flux distribution, $\mathbf{v} = (v_1,v_2,...,v_N)^T$ ($\mbox{g} \cdot \mbox{gDW}^{-1} \cdot \mbox{h}^{-1}$), are automatically satisfied by expressing $\mathbf{v}$ in terms of $K$ vectors $\mathbf{Z}^k$ representing elementary flux modes (EFMs) \cite{Schuster94} that form extremal rays of the so-called `flux cone' (any flux distribution satisfying the stoichiometric constraints of a metabolic reaction network can be represented as a conical combination combination of EFMs, see Figure \ref{fig:1} for illustration) 
\begin{equation*}
\mathbf{v} = \sum_{k=1}^K r_k(\mathbf{m}) \mathbf{Z}^k u_k .
\end{equation*} 
Here $r_k(\mathbf{m})$ denotes the `composite flux' through the $k$th EFM or metabolic pathway and $u_k$ is interpreted as the fraction of total catalytic biomass allocated to it due to the finite resource constraint
\begin{equation}
\label{constraint}
\sum_{k=1}^K u_k = 1, \quad u_k \geq 0 \quad k=1,2,...,K .
\end{equation} 
A detailed interpretation of $r_k(\mathbf{m})$ and $u_k$ in terms of the molecular biology of enzymes catalysing each metabolic reaction can be found in \cite{Tourigny20}, but is not required for this paper. Here the $r_k(\mathbf{m})$ are simply assumed to be non-negative, smooth functions of the slow metabolite concentrations $\mathbf{m}$. The $u_k$ are control variables that must be determined in order to satisfy some optimality criteria and the reduced dynamical system therefore takes the form 
\begin{equation}
\frac{d\mathbf{m}}{dt} = x \sum_{k=1}^K r_k(\mathbf{m}) \mathbf{S} \mathbf{Z}^k u_k,  \quad \frac{dx}{dt} = x \sum_{k=1}^K r_k(\mathbf{m}) \mathbf{c}^T \mathbf{Z}^k u_k    
 \label{reduced}
\end{equation}
where the subscript has been dropped from $\mathbf{S}_{ex}$ for notational convenience.

Extending the above model to include spatiotemporal dynamics proceeds by assuming that multiple interacting metabolic systems of the form (\ref{reduced}) occupy nodes of a directed graph and that slow metabolites are allowed to be diffusively transported across its edges \cite{Barrat08}. Such models typically arise when discretising reaction-diffusion equations and studying diffusively-coupled chemical reactors or biological cells (e.g., \cite{Othmer71,Othmer74,Nakao10}). Abstractly, the graph topology captures spatial organisation in the full spatiotemporal system, and can represent the compartmentalisation of a single cell, tissues or organs in higher-eukaryotes, or spatial structure in the environment or a population of cells. For sake of conciseness, here individual nodes with dynamics (\ref{reduced}) will be taken to represent cells or sub-populations in a larger community and the directed graph that encompasses the entire population will be referred to as the {\em population network}, but application of the model translates to any of the other examples equally well. In the full model consisting of a population network with $\mathcal{N}$ nodes, {\em local} concentrations of slow metabolites and total catalytic biomass at the $i$th node will be denoted by $\mathbf{m}_i$ and $x_i$, respectively, and diffusion of metabolites between nodes is governed by the $(M \times M)$-dimensional diffusion matrices $\mathbf{D}_{ij}$ ($\mbox{h}^{-1}$) such that spatiotemporal dynamics of the entire population take the form
\begin{equation}
\frac{d\mathbf{m}_i}{dt} = x_i \sum_{k=1}^{K_i} r^i_k(\mathbf{m}_i) \mathbf{S}_i \mathbf{Z}^k_i u^i_k + \sum_{j=1}^{\mathcal{N}} \mathbf{D}_{ij} \mathbf{m}_j, \quad \frac{dx_i}{dt} = x_i \sum_{k=1}^{K_i} r^i_k(\mathbf{m}) \mathbf{c}^T_i \mathbf{Z}^k_i u^i_k . 
 \label{full}
\end{equation}
Super- and subscripts $i$ on $\mathbf{S}_i$, $K_i$, $\mathbf{c}_i$, $r^i_k$, and $\mathbf{Z}^k_i$ account for possible differences between the genetic or fixed metabolic traits of cells or sub-populations at each node, and have been included here for full generality. They are not necessary when the same species is assumed to inhabit each node of the population network, in which case $\mathbf{S}_i = \mathbf{S}$, $K_i = K$, $\mathbf{c}_i = \mathbf{c}$, $r^i_k = r_k$, and $\mathbf{Z}^k_i = \mathbf{Z}^k$ for all $i = 1,2,..., \mathcal{N}$. Conversely, superscripts $i$ on the $u^i_k$ are always required because they index the different phenotypic or regulatory traits of cells or sub-populations at each node. Components of the vector $\mathbf{u}^i = (u^i_1,u^i_2,...,u^i_K)^T$ satisfy the constraint (\ref{constraint}) (with $K$ replaced by $K_i$ when considering multiple species), and $u^i_k$ is therefore interpreted as the fraction of local total catalytic biomass at the $i$th node allocated to the metabolic pathway represented by the $k$th EFM. The local concentration of total catalytic biomass, $x_i$, is only able to have an indirect effect on cells or sub-populations at other nodes by contributing to the local production of a slow metabolite, which is then able to diffuse across edges of the population network and participate in metabolic reactions elsewhere.         

The metabolic resource allocation strategy $\mathbf{u}^i$ adopted by the cell or sub-population at node $i$ is dependent on maximising some choice of the general metabolic objective function or performance index $\phi_i$, which in \cite{Tourigny20} was taken to be the local concentration of total catalytic biomass, i.e. $\phi_i = x_i$. This metabolic objective assumes that the species at each node acts independently to maximise $x_i$ with complete disregard for cells or sub-populations at other nodes in the population network. In this case $\mathbf{u}^i$ is an example of an {\em individualist} resource allocation strategy. However, this is not to say there will be no interaction between species using the individualist strategy at different nodes because, as described above, metabolic reactions taking place locally will produce or consume slow metabolites that can diffuse to and participate in reactions at other nodes across the population network, a phenomenon commonly observed in spatially-structured systems that is referred to as metabolic cross-feeding \cite{Nelson05,Belanger11,Semenza08,Lyssiotis17,Schink91,Schink02,Morris13,Song14,Harcombe14,Nadell16,DSouza18,Lin18,Smith19,Evans20}. Metabolic cross-feeding that arises as a consequence of the individualist resource allocation strategy can be interpreted as merely incidental in the sense that it does not necessarily require any evolutionary process to emerge, simply coexistence \cite{Smith19}. It does not fall under the stricter definition of augmented cross-feeding \cite{Smith19} because it comes at no cost (energetic or growth-limiting) to the cell or sub-population producing a metabolite. However, other instances of metabolic cross-feeding may be cooperative in this latter sense: for example, the cell or sub-population at the $i$th node might invest resources into producing a metabolite that could otherwise be invested in maximising its own total catalytic biomass, in favour of species at other nodes. Selection for cooperative metabolic cross-feeding in spatially-structured environments is suggested to be facilitated by various factors such as limited or fluctuating nutrient availability and cell population densities \cite{Bull09,Hoek16,Germerodt16}, in addition to preexistence of incidental metabolic cross-feeding \cite{Pacheco19}. Cooperative metabolic cross-feeding implies that a {\em cooperative} resource allocation strategy $\mathbf{u}^i$ is based on some {\em cooperative metabolic objective} that involves the metabolic performance of cells or sub-populations at other nodes in the population network. The fact that a cooperative metabolic resource allocation strategy may be sub-optimal with respect to an individualistic metabolic objective is equivalent to the statement that metabolite production may be disadvantageous in the absence of the mutualism \cite{Smith19}.  

The cooperative metabolic objective considered here is to maximise the generalised mean of total catalytic biomass $\mathbf{x} = (x_1,x_2,...,x_{\mathcal{N}})^T$ across the population network, which is given by
\begin{equation}
\label{gmean}
M_p(\mathbf{x}) = \left( \frac{1}{\mathcal{N}} \sum_{i=1}^{\mathcal{N}} x^p_i \right)^{1/p}.
\end{equation}   
The generalised mean can also come with a set of weightings $\{ w_i\}_{i=1,2,...,\mathcal{N}}$, one multiplying each $x_i$ and summing to unity, reflecting the fact that different nodes in the population network might not be treated equally. More generally, it is then possible to associate a set of different weights with each node of the network, which would result in a node-dependent metabolic objective function that could potentially better reflect the fact that immediate neighbours of cells or sub-populations are more relevant determinants of the optimal cooperative resource allocation strategy. $M_p(\mathbf{x})$ in fact defines a one-parameter family of cooperative objective functions that includes various analogs of the social welfare functions \cite{Pattanaik08} studied in Welfare Economics and Social Choice Theory as special limiting cases of the real parameter $p$. Interpreted in terms of how the resulting metabolic resource allocation strategy should benefit nodes across the population network, the limiting cases of important relevance are:
\begin{itemize}

\item In the limit $p \to 1$, the metabolic objective becomes maximisation of the arithmetic mean of total catalytic biomass across all nodes in the population network, which is equivalent to the {\em utilitarian} view that resource at the $i$th node should be allocated so as to maximise $x_i$, regardless of the concentrations of total catalytic biomass at others. It turns out that the individualist resource allocation strategy satisfies the utilitarian objective (see Appendix \ref{sec:appendix} for derivation) and, as previously described, any metabolic cross-feeding that arises as a consequence would therefore be considered incidental

\item As $p \to - \infty$, the generalised mean (\ref{gmean}) approaches the minimum function
\begin{equation*}
\lim_{p \to - \infty} M_p(\mathbf{x})  = \min \{ x_1,x_2, ..., x_{\mathcal{N}} \}
\end{equation*}
and the metabolic objective becomes equivalent to the {\em egalitarian} view that resource should be allocated so as to benefit the node with lowest concentration of total catalytic biomass at any given moment in time. Metabolic cross-feeding resulting from a cooperative resource allocation strategy that satisfies the egalitarian objective would tend to direct metabolites away from nodes already with large $x_i$ to those with lower, and might be expected in systems where it is functionally necessary to maintain existence of cells or sub-populations across the entire population network  

\item As $p \to + \infty$, the generalised mean (\ref{gmean}) approaches the maximum function
\begin{equation*}
\lim_{p \to + \infty} M_p(\mathbf{x})  = \max \{ x_1,x_2, ..., x_{\mathcal{N}} \}
\end{equation*}
and the objective becomes equivalent to the {\em elitist} view that resource should be allocated so as to benefit the node with the highest concentration of total catalytic biomass at any given moment in time. In contrast to the egalitarian view, a cooperative resource allocation strategy that satisfies the elitist objective would tend to direct all metabolic cross-feeding towards a single dominant node, with others sacrificing their own growth in its favour. This might be expected in systems where the only matter of importance is to ensure the survival and persistence of a cell or sub-population in at least one location of the population network       

\item In the limit $p \to 0$, most easily evaluated applying L'H\^{o}pital's rule after taking the natural logarithm, the objective reduces to maximisation of the geometric mean of total catalytic biomass across all nodes in the population network, equivalent to maximising the {\em Nash social welfare function} \cite{Nash50} or geometric mean
\begin{equation*}
\lim_{p \to 0} M_p(\mathbf{x})  = (x_1\cdot x_2 \cdots x_{\mathcal{N}} )^{1/\mathcal{N}} ,
\end{equation*}
which favours both increases in overall total catalytic biomass across the population network and inequality-reducing metabolic cross-feeding. In this sense, the Nash social welfare objective may be regarded as a good compromise between the utilitarian and egalitarian objectives, and perhaps provides the best example of a cooperative metabolic resource allocation strategy that is mutually beneficial to all nodes in the population network 

\end{itemize}
Introducing a cooperative objective into the spatiotemporal model of metabolic resource allocation poses a potential challenge, because the form of $\phi_i$ implies that a cell or sub-population at the $i$th node has access to information pertaining to concentrations of total catalytic biomass {\em globally} across the entire population network. Although the cooperative resource allocation strategy $\mathbf{u}^i$ is enacted locally, a biological mechanism must exist for obtaining such a global measurement and this could make interpretation or justification of the model difficult. In the case of cells or sub-populations of microbial species however, cooperative decisions are often based on quorum sensing that is well-understood as a general mechanism for signalling population density in spatially-structured environments (see \cite{Miller01} and references therein). Indeed, quorum sensing is known to regulate a wide range of metabolic activity including metabolic cross-feeding \cite{An14,Goo15,Davenport15} and is therefore the clear candidate for communicating concentrations of total catalytic biomass between nodes in the population network. For example, a quorum sensing molecule may serve as an indicator for population density and regulate (through gene expression or post-transcriptional mechanisms) the relative activity of metabolic pathways in response to increased cell crowding \cite{An14,Goo15}. Moreover, integration of metabolism and quorum sensing has recently been shown to govern cooperative behaviour in populations of {\em Pseudomonas aeruginosa} and other organisms \cite{Boyle15,Boyle17}. The way that quorum sensing manifests heterogeneously \cite{Grote15} in response to environmental fluctuations and uncertainty \cite{Popat15} also speaks to the cooperative maximum entropy control introduced in the next section. It is not immediately clear whether such a mechanism exists for multi-cellular organisms, but almost certainly higher-eukaryotes can exploit the circulation of hormones and other signalling molecules for communication between spatially-separated systems.   

\section{Cooperative maximum entropy control} 
\label{sec:control}
In \cite{Tourigny20}, a metabolic resource allocation strategy based on the maximum entropy principle was motivated from the perspective of bet-hedging in a temporally-fluctuating environment \cite{Seger87,Perkins09,Ackermann15,Granados17} because, up to a constant factor, entropy uniquely satisfies the accepted axioms for an uncertainty measure \cite{Shannon48}. The maximum entropy distribution is therefore uniquely determined as the one consistent with known constraints (e.g., maximising expected return-on-investment given current environmental conditions) that expresses maximum uncertainty with respect to everything else (e.g., future environmental fluctuations) \cite{Jaynes57,Shore80}. This could also involve aspects of regulation that have not been otherwise shaped by natural selection. Metabolic resource allocation based on the maximum entropy principle and its possible implementation via population heterogeneity \cite{Ridden15,DeMartino17,Fernandez19} is paralleled by the statement that optimal phenotype-switching strategies must accommodate entropies of temporally-fluctuating environments \cite{Kussell05}, and could readily be extended to include costs associated with switching between different metabolic pathways. However, in spatially-structured systems further uncertainty arises due to the physical limit on knowledge that cells or sub-populations can access beyond their immediate local (spatial) environment, and this lack of detailed global information makes cooperative behaviour difficult to understand without an extension of the maximum entropy bet-hedging strategy. In this section, a cooperative maximum entropy control law is derived for a metabolic resource allocation strategy $\mathbf{u}^i$ based on the cooperative metabolic objective of maximising the generalised mean (\ref{gmean}). This results in a family of control laws that explain the beneficial properties of metabolic cross-feeding, for various special limiting cases of the parameter $p$. 

Assuming the genetic or fixed traits of species are identical at all nodes in the population network, without loss of generality consider the $i$th whose dynamics are of the form (\ref{full}) and resource allocation strategy $\mathbf{u}^i$ is determined by the cooperative metabolic objective function $\phi_i = M_p(\mathbf{x})$. By analogy with \cite{Tourigny20}, this cooperative metabolic objective function is combined with the maximum entropy principle such that $\mathbf{u}^i$ (the maximum entropy control) takes the form of a Boltzmann distribution. In this Boltzmann distribution, `energy' of the $k$th metabolic pathway or EFM is replaced by {\em effective return-on-investment}, i.e., the expected contribution to the cooperative metabolic objective function if all resource were to be exclusively allocated to that EFM, and a `temperature' factor, $\sigma$, controls the relative spread across EFMs according to their effective return-on-investment. The effective return-on-investment for the $k$th EFM at the $i$th node is
\begin{equation}
\label{eroi}
\mathcal{R}_{\Delta t}^{k,i} = \mathbf{q}_i^T \mathbf{e}^{\mathbf{A} \Delta t} \mathbf{B}^k_i
\end{equation}   
where $\mathbf{q}_i = \partial \phi_i(\mathbf{X}(t))/\partial \mathbf{X}$ is the gradient of $\phi_i$ evaluated at $\mathbf{X}(t)= (\mathbf{m}_1,x_1, ..., \mathbf{m}_{\mathcal{N}},x_{\mathcal{N}})^T$, $\mathbf{e}^{\mathbf{A} \Delta t}$ the matrix exponential of $\Delta t$ times the full Jacobian matrix $\mathbf{A}$ of system (\ref{full}) evaluated at $\mathbf{X}(t)$ and reference controls $\mathbf{u}^i_0$ ($i=1,2,...,\mathcal{N}$ ), and $\mathbf{B}^k_i$ is the derivative of (\ref{full}) with respect to $u^i_k$ evaluated at $(\mathbf{m}_i, x_i)^T$. The cooperative maximum entropy control for the $i$th node then takes the form
\begin{equation}
\label{control}
u^i_k(t) = \frac{1}{Q_i}\exp \left( \mathcal{R}_{\Delta t}^{k,i} /\sigma  \right) , \quad k=1,2,...,K
\end{equation}  
where $Q_i = \sum_{k=1}^K \exp ( \mathcal{R}_{\Delta t}^{k,i} /\sigma)$ is a normalisation factor (the partition function) and $\sigma$ a positive parameter governing the spread of resource among EFMs at the $i$th node. As described previously \cite{Tourigny20}, the control (\ref{control}) collapses to the DFBA policy \cite{Mahadevan02} in the limit $\sigma \to 0$, since then all resource is allocated exclusively to the EFM with the greatest effective return-on-investment (\ref{eroi}). Conversely, $u^i_k \to 1/K$ $(\forall k = 1,2,..., K)$ as $\sigma$ grows so that resource is partitioned equally among all EFMs in the limit $\sigma \to \infty$. This latter scenario is equivalent to the unregulated macroscopic bioreaction models of Provost and Bastin \cite{Provost04,Provost06}. Thus, $\sigma$ captures the bet-hedging nature \cite{Seger87,Perkins09} of dynamic metabolic resource allocation, which could be implemented by exploiting heterogeneity at the population level \cite{Ackermann15,DeMartino17,Fernandez19}, distributed regulation within each cell \cite{Granados17}, or a combination of both mechanisms enacted simultaneously. The maximum entropy control further implies that all EFMs (even those with with zero or negative effective return-on-investment) will always be allocated a non-zero fraction of metabolic resource, which distinguishes it from previous resource allocation strategies including those of Young and Ramkrishna \cite{Young07}. Equipped only with knowledge about current environmental conditions, allocating a small fraction of resource to EFMs not contributing directly to growth is a bet-hedging strategy not considered wasteful because there is always a small probability that these pathways will have a benefit in the future \cite{OBrien16,Mori17,Tourigny20}. Appearance of the Jacobian matrix $\mathbf{A}=\mathbf{A}(\mathbf{X})$ in the effective return-on-investment (\ref{eroi}) is equivalent to the biological statement that regulatory decisions may also take into consideration effects the control action will have on the environment, and therefore individual EFMs not contributing directly to growth can receive greater (or less) investment should they involve consumption or production of metabolites that make the environment more (or less) favourable. Consequently \cite{Tourigny20}, the maximum entropy principle has been used to explain the accumulation of metabolic reserves under nutrient-limiting conditions \cite{Lillie80,Francios01}. The cooperative maximum entropy control (\ref{control}) likewise describes the phenomenon of metabolic cross-feeding, analogous to allocation of resources to metabolic pathways that do not contribute directly to the instantaneous growth rate of an individual cell or sub-population, as described below. Stated in terms of cooperative behaviour, this means investment in metabolic pathways that benefit other nodes in the population network at an apparent cost to the $i$th.    
 
Evaluation of (\ref{control}) depends on a choice of $\Delta t = 0$, resulting in the {\em greedy} cooperative maximum entropy control law, or $\Delta t >0$, resulting in the {\em temporal} cooperative maximum entropy control law \cite{Young07,Tourigny20}. In effect, the difference between these two control laws pertains to how much of the future consequences of a regulatory action or changes in environmental conditions are taken into consideration when deriving the corresponding metabolic resource allocation strategy. Only the greedy cooperative maximum entropy control law ($\Delta t = 0$, only instantaneous effects considered) will be studied in this paper. The distinction between greedy and temporal control laws is more than one of computational efficiency, because inclusion of $\mathbf{A}$ in the temporal cooperative maximum entropy control implies that the cell or sub-population at the $i$th node has a predictive capacity not only for future events, which could be anticipated using some form of regulation, but also specific information on the level of metabolic activity at nodes elsewhere in the population network. Obtaining the latter may be especially difficult to justify on biological grounds. However, although the greedy cooperative maximum entropy control also takes into account both local and global information, the only global information required is an overall measure of the total catalytic biomass across the entire population network that comes from $\mathbf{q}_i$ in (\ref{eroi}). Excluding the utilitarian objective with $p=1$ that yields an individualist maximum entropy control identical to \cite{Tourigny20}, this is most intuitively seen by setting $\Delta t = 0$ in (\ref{eroi}) and expressing the greedy effective return-on-investment as (see Appendix \ref{sec:appendix} for full derivation)
\begin{equation}
\label{geroi}
\mathcal{R}_0^{k,i} = \frac{x^p_i}{x^p_i + y^p} \cdot M_p(\mathbf{x}) \cdot R^k_0(\mathbf{m}_i) .
\end{equation}         
Here $R^k_0(\mathbf{m}) = r_k(\mathbf{m}) \mathbf{c}^T \mathbf{Z}^k$ is the zeroth-order return-on-investment for the $k$th EFM described in \cite{Tourigny20}, and notation $y$ has been introduced for the generalised sum
\begin{equation*}
y = \left( \sum_{j\neq i}^{\mathcal{N}} x_j^p \right)^{1/p}
\end{equation*}
of total catalytic biomass across the population network not involving the $i$th node. The terminology {\em effective} return-on-investment was coined in \cite{Tourigny20} because, for the greedy individualist maximum entropy control considered there, each $u_k$ came with a common factor of $x$ multiplying $R^k_0(\mathbf{m})/\sigma$ in the exponent. Similarly, for the cooperative control (\ref{control}) considered here, the greedy effective return-on-investment (\ref{geroi}) comes with $R^k_0(\mathbf{m}_i)$ multiplied by two factors: the first, $M_p(\mathbf{x})$, is common to all nodes and EFMs in the population network (i.e., $M_p(\mathbf{x})$ appears in $u^i_k$ for all $i=1,2,...,\mathcal{N}$, $k=1,2,...,K$) and therefore plays a role analogous to $x$ in \cite{Tourigny20}, since resource will become further concentrated on the EFMs with largest $R^k_0(\mathbf{m})$ as the value of the objective function $\phi_i = M_p(\mathbf{x})$ increases. The second, the sigmoid function $x^p_i/(x^p_i + y^p)$, appears as a common factor in the exponent of $u^i_k$ for the $i$th node (i.e., $x^p_i/(x^p_i + y^p)$ appears in $u^i_k$ for all $k=1,2,...,K$), but will take on different values at different nodes in accordance with how the local concentration of total catalytic biomass, $x_i$, compares to the global ensemble, $y$, as measured across the remainder of the population network. In this sense the greedy cooperative maximum entropy control $\mathbf{u}^i$ requires both local and global information, but does not depend on the cell or sub-population at the $i$th node having specific knowledge of the metabolic activity or environmental composition at any location other than its own. Instead, only the overall measures of global population network performance provided by $y$ and $M_p(\mathbf{x})$ must be accessible for enacting optimal regulatory decisions locally. As explained in Section \ref{sec:model}, an obvious candidate for communicating this information is quorum sensing, which is well-understood as a generic biological mechanism for signalling population density and regulating metabolism in microbial systems \cite{Miller01,An14,Goo15,Davenport15,Boyle15,Boyle17}. 

\begin{figure}
    \caption{Plots of the sigmoid $x^p_i/(x^p_i + y^p)$ multiplying $M_p(\mathbf{x}) \cdot R^k_0(\mathbf{m}_i)$ as a function of $x_i$ in the greedy effective return-on-investment (\ref{geroi}) for low (egalitarian regime) and high (elitist regime) values of $p$. The value of the generalised sum $y$ is displayed on the $x$-axis. In the limit $p \to 0$, the sigmoid converges to the flat line with intercept $1/2$ (represented by dashed line in panels), while for $p=1$ the denominator of the sigmoid cancels the arithmetic mean $M_1(\mathbf{x})$ and (\ref{geroi}) reduces to the greedy effective return-on-investment for the individualistic maximum entropy control \cite{Tourigny20}. \label{fig:fig1}}
    \centering
     \begin{subfigure}[t]{0.45\textwidth}
     \centering
     \includegraphics[width=\linewidth]{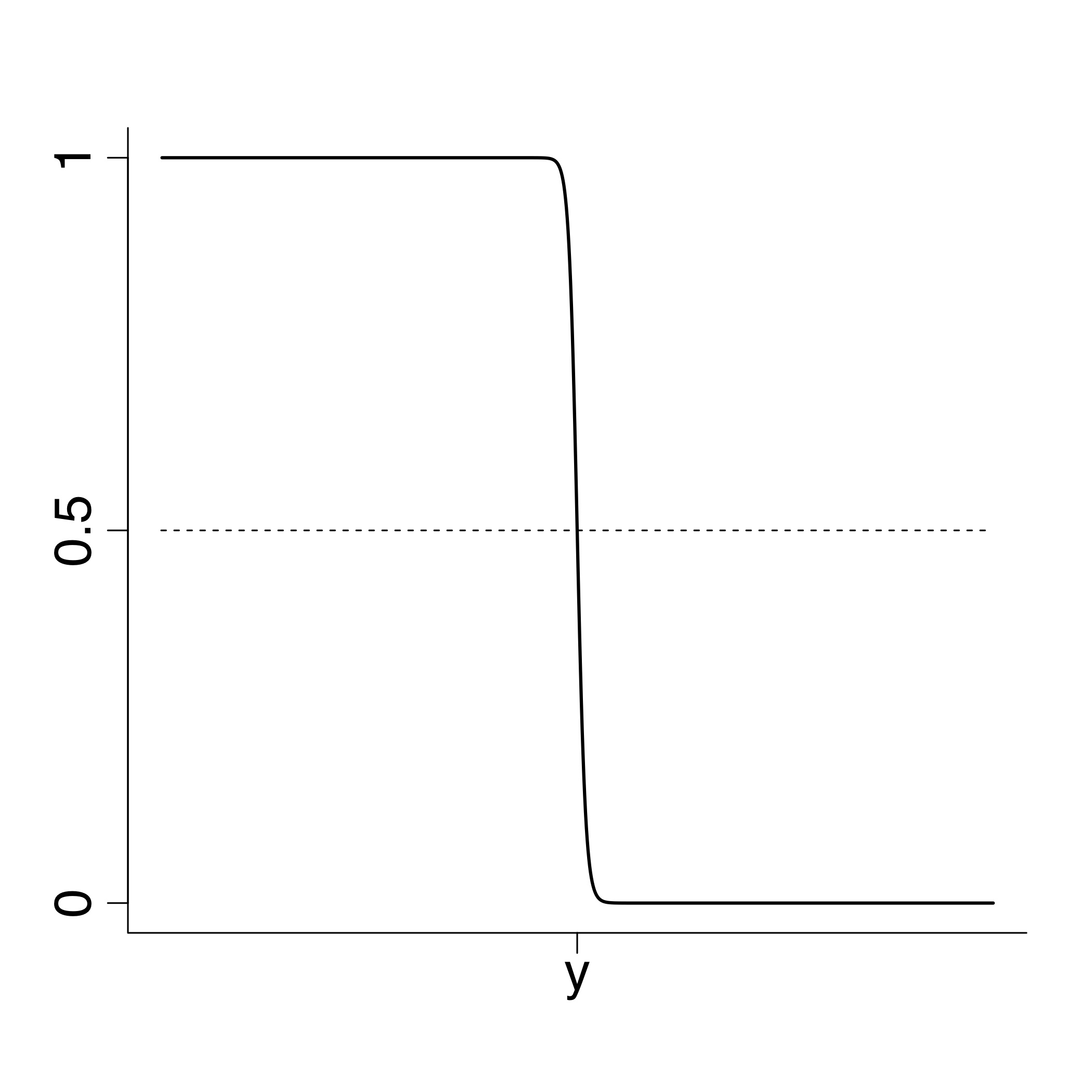} 
     \caption{Plot of the sigmoid for $p=-100$.} \label{fig:2a}
    \end{subfigure}
    \hfill
    \begin{subfigure}[t]{0.45\textwidth}
        \centering
        \includegraphics[width=\linewidth]{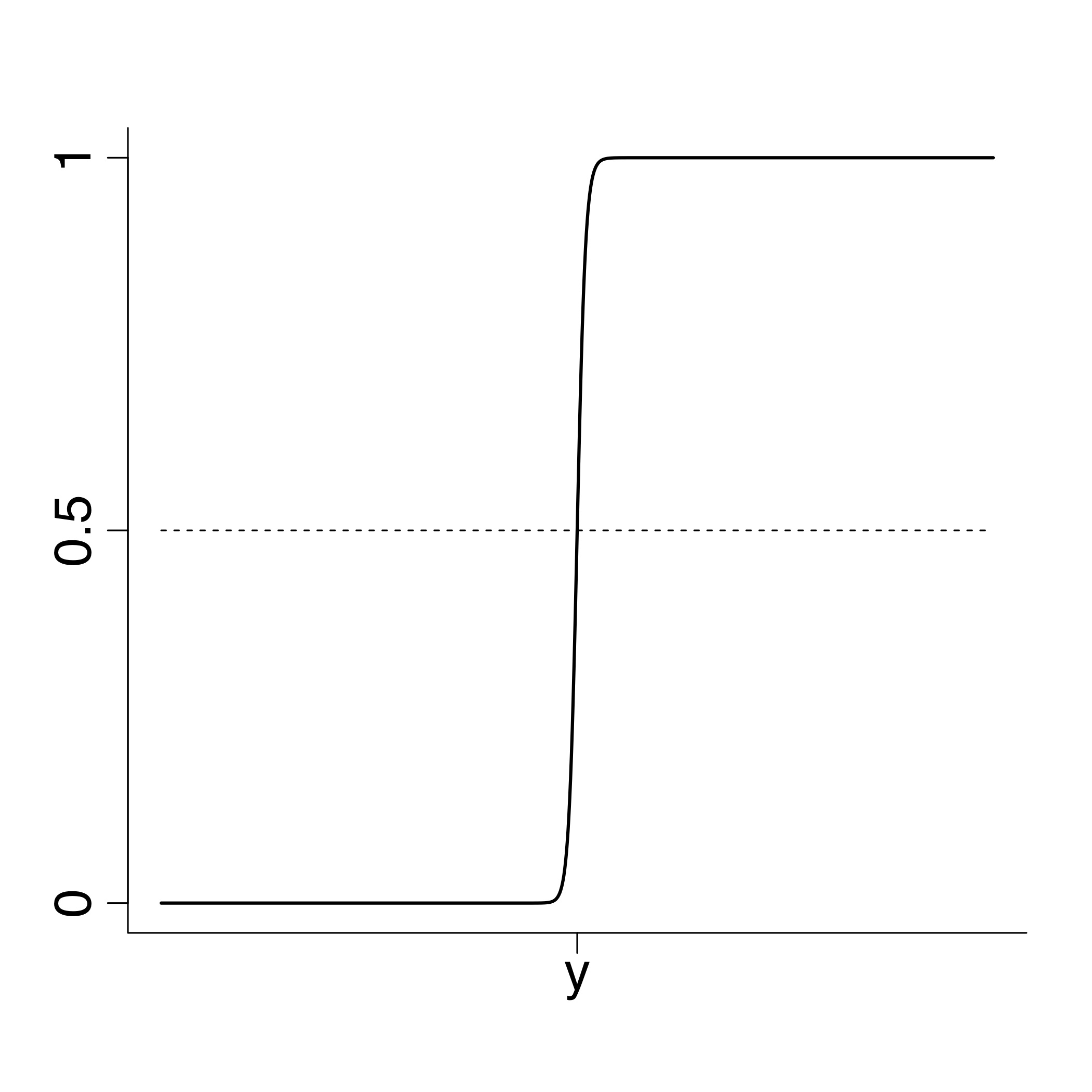} 
        \caption{Plot of the sigmoid for $p=100$.} \label{fig:2b}
    \end{subfigure}
\end{figure}

In Figure \ref{fig:fig1}, the sigmoid function in (\ref{geroi}) is displayed for low and high values of $p$ where $M_p(\mathbf{x})$ approaches the egalitarian and elitist objective functions, respectively, and illustrates how it approximates a Heaviside step function with change point at $y$. For the egalitarian regime with $p << 0$ the sigmoid $x^p_i/(x^p_i + y^p)$ is close to zero when $x_i > y$, but close to unity when $x_i < y$. This implies that the greedy effective return-on-investment (\ref{geroi}) is relatively small when $x_i > y$, which indicates a larger spread of resource across EFMs at the $i$th node regardless of their zeroth-order return-on-investments $R^k_0(\mathbf{m}_i)$. Consequently, cells or sub-populations whose local concentration of total catalytic biomass is large relative to that of the remaining population will tend to distribute resource more indiscriminately among EFMs, which facilitates metabolic cross-feeding by increasing investment in metabolic pathways that contribute to the production of a metabolite rather than increasing their own local growth rate. Conversely, the greedy effective return-on-investment (\ref{geroi}) will be large at nodes with a local concentration of total catalytic biomass that is small relative to that of the remaining population (i.e., when $x_i < y$). Cells or sub-populations at these nodes will instead concentrate their resources on EFMs with large $R^k_0(\mathbf{m})$ in order to maximise their own local concentration of total catalytic biomass directly. The combined effect of this type of cooperative dynamic behaviour will be to increase the total catalytic biomass of the cell or sub-population that has the lowest local concentration across the population network at any given moment in time. When $p >> 0$, corresponding to the elitist regime, the situation is reversed in that a node with $x_i > y$ will invest greater resource into metabolic pathways maximising $x_i$, while the spread of resource will be greater and favour production of metabolites for metabolic cross-feeding if $x_i <y$. The combined effect of the cooperative dynamic behaviour that emerges in this case will be to increase the total catalytic biomass of the cell or sub-population with the highest local concentration across the population network at any given moment in time. Between these two extremes is the Nash regime corresponding to $p=0$, where metabolic cross-feeding is instead controlled uniformly across all nodes in the population network because the common factor of $\frac{1}{2} \sqrt{x_1x_2 \cdots x_{\mathcal{N}}}$ multiples $R^k_0(\mathbf{m}_i)/\sigma$ in the exponent of all $u^i_k$, and therefore plays an equivalent role to $x_i$ in the individualist greedy maximum entropy control. This reflects the compromise between the utilitarian and egalitarian objectives: the cell or sub-population at each node takes into consideration overall population performance when locally regulating the spread of resource across metabolic pathways, and all nodes will tend to favour metabolic cross-feeding (larger spread of resource) when the value of the Nash objective function is low.  

The greedy cooperative maximum entropy control law captures several of the relationships between metabolic cross-feeding, nutrient limitation, growth rate, and population density that have been studied previously. Population density has been suggested to determine the conditions that favour metabolic cross-feeding \cite{Bull09,Estrela10}, and the same conclusion applies to the cooperative resource allocation strategy considered here if one uses total catalytic biomass to infer population density, which in turn regulates the spread of resource across metabolic pathways as described above. Analogous to the inverse correlation between growth rate and the production of storage metabolites that exists for the individualistic maximum entropy control \cite{Tourigny20}, investment of local resource into pathways producing metabolites for cross-feeding implies a reduction in the local growth rate of the cell or sub-population at the $i$th node. Reduction of local growth rate is in turn another form of cooperation that provides indirect benefit to neighbouring cells or sub-populations by reducing competition for limited extracellular metabolites \cite{Gardner07}; in this sense, the bet-hedging nature of any resource allocation strategy based on the maximum entropy principle can be considered cooperative behaviour in its own right \cite{Perkins09}. Recently, Germerodt et al. \cite{Germerodt16} found that spatially-structured environments with fluctuating nutrient availability tend to favour cooperative metabolic cross-feeding only when metabolites drop below a certain level, which can be explained by appearance of the zeroth-order return-on-investment $R^k_0(\mathbf{m})$ in the greedy return-on-investment (\ref{geroi}). As for the individualistic maximum entropy control \cite{Tourigny20}, the greedy cooperative maximum entropy control takes into consideration both the nutrient composition of the environment and concentrations of total catalytic biomass (local versus global), resulting in a metabolic resource allocation strategy determined by the product  (\ref{geroi}) rather than sum of these quantities. Finally, it is important to highlight that, for the cooperative metabolic objective function (\ref{gmean}) considered here, the maximum entropy control law (\ref{control}) is the only metabolic resource allocation strategy considered to date that will describe such cooperative behaviour in a biologically-realistic setting with $\Delta t = 0$. This follows from the observation that a common factor multiplies $R^k_0(\mathbf{m}_i)$ for all $k=1,2,...,K$ in the greedy return-on-investment (\ref{geroi}), and therefore a DFBA control law obtained from $\mathbf{u}^i$ in the limit $\sigma \to 0$ would result in all resource being allocated exclusively to the EFM with largest zeroth-order return-on-investment, regardless of the local concentration of total catalytic biomass at the $i$th node. Similarly, the same multiplicative factor cancels out in the greedy control law derived by Young and Ramkrishna \cite{Young07}, again resulting in a resource allocation strategy that only takes into consideration the $R^k_0(\mathbf{m}_i)$. In their current form, failure of these alternative control laws to account for cooperative metabolic behaviour further supports the original proposal \cite{Tourigny20} of a model for dynamic metabolic resource allocation based on the maximum entropy principle. This is exemplified by the application described in the next section.   
         
\section{Application to microbial biofilms and colonies}
\label{sec:example}
This section introduces a model for cooperative metabolic resource allocation in a microbial community consisting of a single species, such as a population of genetically-identical bacteria or yeast growing as a biofilm or colony. These systems often come with a certain degree of spatial organisation because access to nutrients can be limited to particular locations within the total population (e.g., at the extremities of a biofilm or base of a colony) and, consequently, heterogenous phenotypic traits may be adopted by cells or sub-populations occupying different spatial domains. Some of the best-characterised experimental systems are biofilms formed by the pathogenic bacteria {\em P. aeruginosa} \cite{Mulcahy14,Rasamiravaka15}, which have motivated several spatiotemporal modelling frameworks that can be used to understand metabolic cross-feeding (e.g., \cite{Biggs13,Chen16,Zhang18}). {\em P. aeruginosa} is of even greater relevance here, because as a species they are one of the prime examples for which a clear role has been established for quorum sensing in the regulation of cooperative metabolic behaviour \cite{Davenport15,Boyle15,Boyle17}, possibly including lactate-based metabolic cross-feeding \cite{Lin18}. Other experimental systems to which the model applies equally well, but where a detailed understanding of quorum sensing is comparatively lacking, are colonies of {\em Escherichia coli} that display a form of acetate-based metabolic cross-feeding \cite{Cole15,Peterson17,Wolfsberg18}, and colonies of yeast that exploit similar lactate- \cite{Cap12} or trehalose-based \cite{Varahan19} mechanisms. 

\begin{figure}
     \centering
     \includegraphics[width=\linewidth]{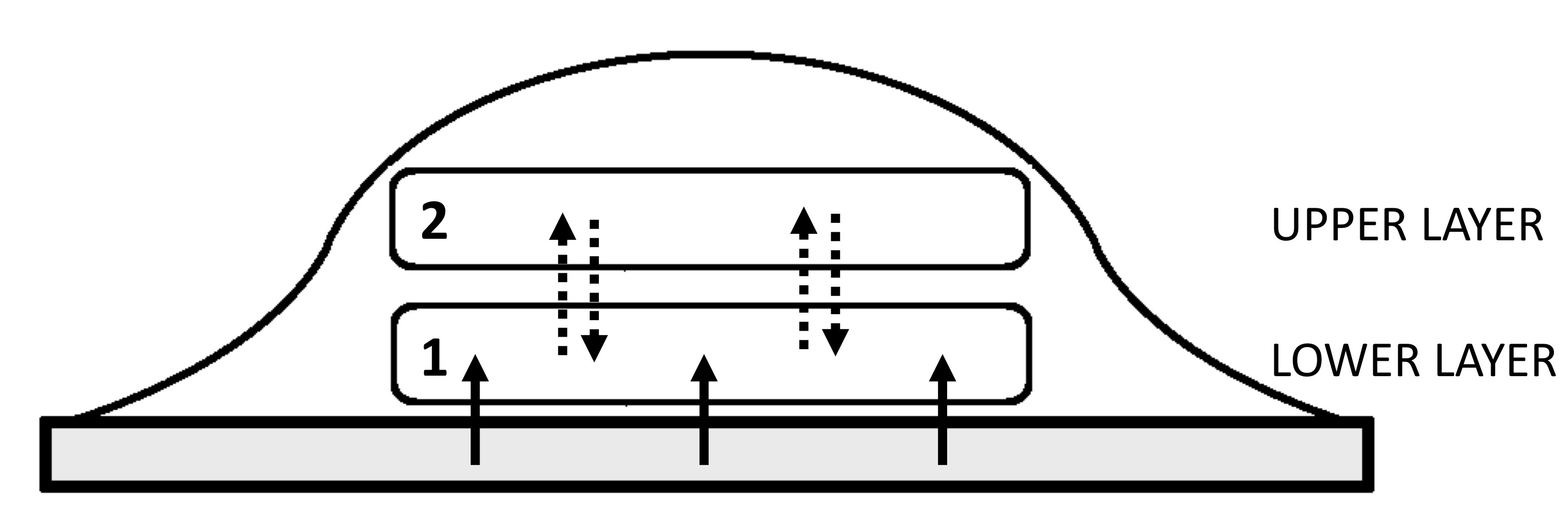} 
       \caption{Cartoon illustration of the population network from the simple model in Section \ref{sec:example} as a microbial colony growing on an agar substrate containing glucose as a limiting nutrient. Spatial structure of the colony can be approximated as a lower and upper layer, represented by nodes $i=1$ and $i=2$ in the population network, respectively. Only cells in the lower layer that is in contact with the agar substrate have access to glucose (represented by solid arrows), while oxygen and the fermentation product (represented by dashed arrows) are free to diffuse between both layers. \label{fig:3}}
\end{figure}

The model constructed following the framework outlined in previous sections is highly simplified in that the population network consists of just two nodes representing two distinct spatial environments in the colony or biofilm. These are intended to capture either the interior and exterior regions of a three-dimensional biofilm, or the upper and base layers of a raised colony growing on an agar substrate (Figure \ref{fig:3}). The metabolic reaction network of the microbial species at each node is taken to be a simplified model of central carbon metabolism (Figure \ref{fig:1a} and Example 1 in \cite{Tourigny20}, see also \cite{Moller18}), resulting in local concentrations of glucose ($G_i$), oxygen ($O_i$), a fermentation product ($P_i$), and total catalytic biomass ($x_i$) ($i =1,2$ in each case) as kinetic variables. The full dynamical system expressed in terms of the three EFMs (Figure \ref{fig:1b} and Example 1 in \cite{Tourigny20}) corresponding to the following three metabolic pathways: glucose fermentation ($k=1$); oxidation of the fermentation product ($k=2$); and oxidation of glucose ($k=3$), is given by the equations
\begin{eqnarray*}
\frac{dG_1}{dt} &=&  -\frac{1}{2}[ r_1(G_1)u^1_1 + r_3(G_1,O_1)u^1_3] x_1 + D_G(G_2 - G_1)\\
\frac{dO_1}{dt} &=&   kLa(O^* - O_1) -[r_2(O_1,P_1)u^1_2 + r_3(G_1,O_1)u^1_3] x_1 + D_O(O_2 - O_1) \\
\frac{dP_1}{dt} &=& [r_1(G_1)u^1_1 -  r_2(O_1,P_1)u^1_2]x_1 + D_P(P_2 - P_1)   \\
\frac{dx_1}{dt} &=& [\frac{1}{2}c_1r_1(G_1)u^1_1  + c_3  r_2(O_1,P_1)u^1_2 + \frac{1}{2}(c_1 + 2c_3)  r_3(G_1,O_1)u^1_3 ]x_1
\end{eqnarray*}
\begin{eqnarray*}
\frac{dG_2}{dt} &=&  -\frac{1}{2}[ r_1(G_2)u^2_1 + r_3(G_2,O_2)u^2_3] x_2 + D_G(G_1 - G_2)\\
\frac{dO_2}{dt} &=&   kLa(O^* - O_2) -[r_2(O_2,P_2)u^2_2 + r_3(G_2,O_2)u^2_3] x_2 + D_O(O_1 - O_2) \\
\frac{dP_2}{dt} &=& [r_1(G_2)u^2_1 -  r_2(O_2,P_2)u^2_2]x_2 + D_P(P_1 - P_2)   \\
\frac{dx_2}{dt} &=& [\frac{1}{2}c_1r_1(G_2)u^2_1  + c_3  r_2(O_2,P_2)u^2_2 + \frac{1}{2}(c_1 + 2c_3)  r_3(G_2,O_2)u^2_3 ]x_2
\end{eqnarray*}
where volumetric mass transfer coefficient $k_La$ and dissolved oxygen solubility limit $O^*$ have been introduced to model oxygen supply, and $D_G,D_O,D_P$ are the diffusion coefficients of extracellular glucose, oxygen, and the fermentation product, respectively. As described in \cite{Tourigny20}, the $r_k$ are approximated by Michaelis-Menten kinetics according to the metabolites whose uptake fluxes are in the support of each EFM, such that
\begin{eqnarray*}
r_1(G_i) &=& V^{max}_1 \frac{G_i}{K_1 + G_i} \\ 
r_2(O_i,P_i) &=& V^{max}_2 \frac{P_i}{K_2 + P_i}\frac{O_i}{K_{O,2} + O_i} \\
r_3(G_i,O_i) &=& V^{max}_3 \frac{G_i}{K_3 + G_i}\frac{O_i}{K_{O,3} + O_i}
\end{eqnarray*}
and consequently the zeroth-order return-on-investments are
\begin{equation*}
R^1_0(G_i) =  \frac{1}{2}c_1 r_1(G_i), \quad R^2_0(O_i,P_i) = c_3 r_2(O_i,P_i) , \quad R^3_0(G_i,O_i) = \frac{1}{2}(c_1 + 2c_3) r_3(G_i,O_i)
\end{equation*}  
for $i = 1,2$. The greedy cooperative maximum entropy control law for determining each $u^i_k$ is then
\begin{equation*}
u^i_k = \frac{1}{Q_i} \exp \left( \frac{1}{\sigma}   \frac{x^p_i}{x^p_i + y^p} \cdot M_p(\mathbf{x}) \cdot R^k_0(\mathbf{m}_i) \right) , \quad y = x_{j \neq i}
\end{equation*} 
with $\mathbf{m}_i = (G_i,O_i,P_i)^T$ and where $Q_i$ is the partition function used to normalise $\mathbf{u}^i$ such that $\sum_{k=1}^3 u^i_k = 1$.

Numerical simulations of this system were performed using custom-built software based on SUNDIALS solvers \cite{Hindmarsh05}, employing the parameter values and initial conditions displayed in Table \ref{tab:table1}. For these simulations, initial conditions and parameter values are chosen to model the effect of having no glucose available to the cell or sub-population at node $i=2$ (e.g., this node might correspond to the upper layer of a colony growing on an agar substrate as in Figure \ref{fig:3}), but a single (exhaustible) source of glucose available to the cell or sub-population at node $i=1$. Remaining kinetic parameter values are generic as in \cite{Tourigny20} such that no attempt has been made to fit them to experimental data although are comparable in magnitude to those from a related model \cite{Jones99}. Predictions of the model should therefore be treated as purely qualitative and are intended to capture effects of the maximum entropy control law without augmentation based on additional biological knowledge, such as pathway costs that can enhance the capacity for overflow metabolism \cite{Moller18,Basan15}. Values for the diffusion coefficients reflect the fact that, while oxygen and the fermentation product are free to diffuse between nodes in the population network, glucose may not (i.e., $D_G = 0.0$ $\mbox{h}^{-1}$). In particular, the concentration of glucose at the second node remains  identically zero across the entire simulation, $G_2(0) \equiv 0$, so that the cell or sub-population at node $i=2$ is obligatorily dependent on metabolic cross-feeding, and can only grow using the fermentation product produced from the cell or subpopulation at node $i=1$. However, for the parameter values in Table \ref{tab:table1}, metabolic pathway $k=3$ (glucose oxidation) has the largest zeroth-order return-on-investment when the local concentrations of glucose and oxygen are sufficiently high. Thus, biologically, the cell or sub-population at node $i=1$ must chose between allocating greater resource towards this pathway so as to increase its own local growth rate, or assisting the cell or sub-population at node $i=2$ to increase its local concentration of total catalytic biomass by allocating a bigger fraction of resource to metabolic pathway $k=1$ (glucose fermentation). In the model, these choices are determined by different values of the parameter $p$ in the cooperative metabolic objective function (\ref{gmean}). 

\begin{figure}
     \centering
     \includegraphics[width=\linewidth]{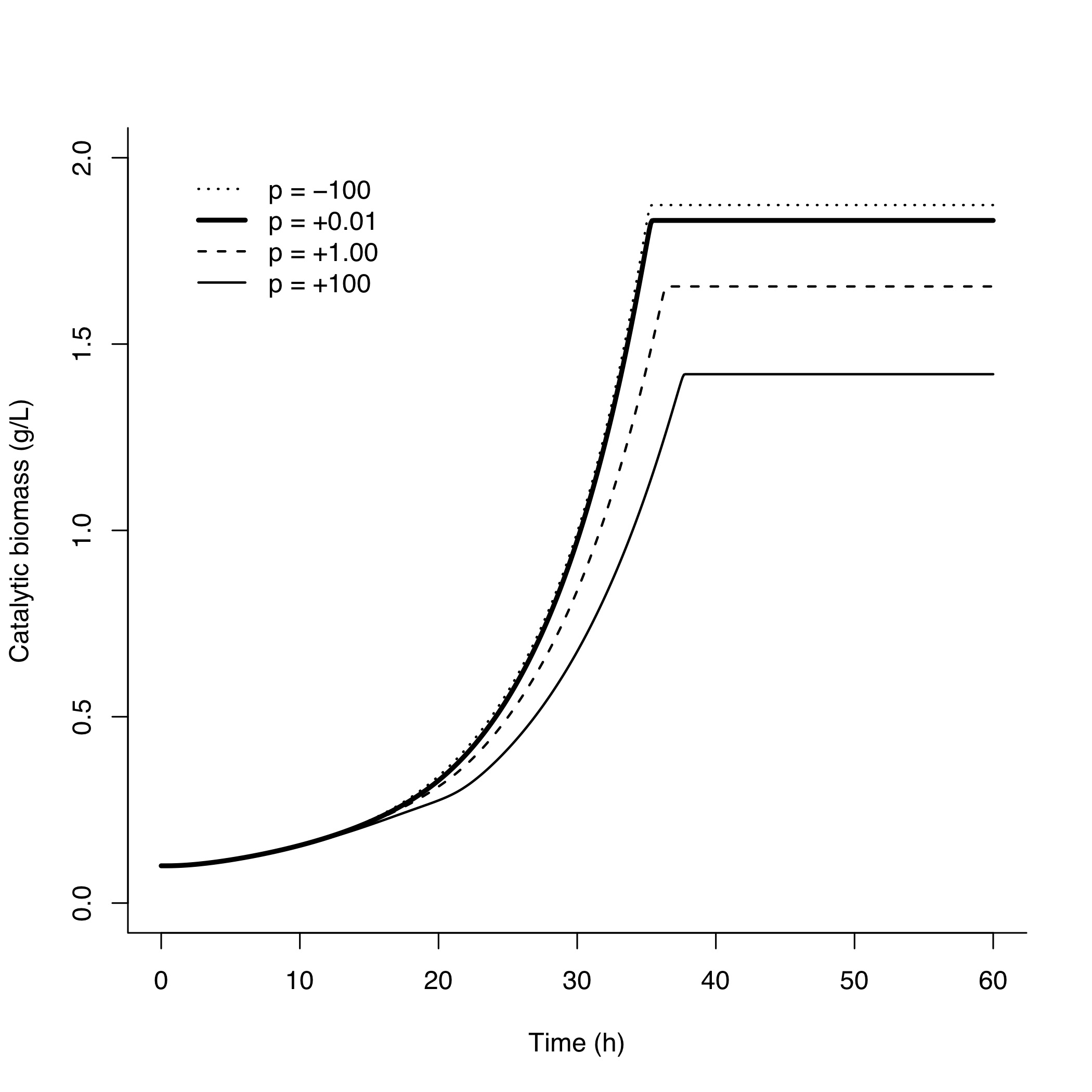} 
         \caption{Trajectories for the local concentration of total catalytic biomass $x_2$ obtained from simulation of the model described in Section \ref{sec:example}, plotted for relevant values of $p$ corresponding to the individualistic regime  ($p=1$), the egalitarian regime ($p=-100$), the elitist regime ($p=100$), and the Nash regime ($p=0.01$). \label{fig:4}}
\end{figure}

Results of numerical simulations are displayed in Figure \ref{fig:4} for various values of $p$, approximating the individualistic, egalitarian, elitist, and Nash regimes of cooperation, respectively. In the individualistic regime, the cell or sub-population at node $i=2$ is still able to increase $x_2$ at a reasonable rate due to incidental metabolic cross-feeding that arises as a consequence of the bet-hedging nature of the individualist maximum entropy control, which models a form of diauxic growth for the cell or sub-population at node $i=1$. When increasing the value of $p$ towards the the elitist regime however, both the local growth rate and final local concentration of total catalytic biomass at node $i=2$ decrease substantially. This is because the cooperative maximum entropy control disfavours metabolic cross-feeding at the dominant node in the elitist regime, and consequently there is less fermentation product available for the cell or sub-population at node $i=2$ to grow. At the other extreme, the egalitarian regime, both the local growth rate and final local concentration of total catalytic biomass at node $i=2$ is substantially higher than in the individualist regime. This is because the cooperative maximum entropy control favours metabolic cross-feeding at the dominant node in the egalitarian regime, and consequently there is more fermentation product available for the cell or sub-population at node $i=2$ to grow. Curiously, for this model, relative to the differences in growth trajectories at node $i=2$ under the individualistic, egalitarian, and elitist metabolic objectives, respectively, there is barely any noticeable difference between the growth trajectories corresponding to Nash and egalitarian control laws. Even though the final local concentration of total catalytic biomass at node $i=2$ achieved is slightly higher under the egalitarian cooperative maximum entropy control, the exponential phase of the $x_2$ trajectory is virtually indistinguishable from that obtained in the Nash regime. This observation could indicate that the Nash regime of the cooperative metabolic objective (\ref{gmean}) has some special biological significance perhaps attributed to its compromising egalitarian-individualist nature. Whether this is a general feature or instead only true in special cases like that presented here remains to be tested by extending the model to other systems.   

\begin{table}[h!]
  \begin{center}
    \caption{Values for parameters and initial conditions used in all simulations. Values for $p$ are reported in Figure \ref{fig:4}.}
    \label{tab:table1}
    \begin{tabular}{l r c l r}
      \textbf{Parameters} & \textbf{Value}  & & \textbf{Initial conditions} & \textbf{Value} \\
      \hline
      $V_1^{max},V_2^{max},V_3^{max}$ & $1.0$ $\mbox{h}^{-1}$ & & $x_1(0), x_2(0)$ & $0.1$ $\mbox{g}\cdot \mbox{L}^{-1}$ \\
      $c_1$ & $0.02$ $\mbox{g}\cdot \mbox{g}^{-1}$ & & $P_1(0), P_2(0)$ & $0.0$ $\mbox{g}\cdot \mbox{L}^{-1}$\\
      $c_3$ & $0.34$ $\mbox{g}\cdot \mbox{g}^{-1}$ & & $O_1(0), O_2(0)$ & $0.0001$ $\mbox{g}\cdot \mbox{L}^{-1}$ \\
      $K_1,K_2,K_3$ & $0.01$ $\mbox{g}\cdot \mbox{L}^{-1}$ & & $G_1(0)$ & $10.0$ $\mbox{g}\cdot \mbox{L}^{-1}$\\
      $K_{O,2},K_{O,3}$ & $0.001$ $\mbox{g}\cdot \mbox{L}^{-1}$ & & $G_2(0)$ & $0.0$ $\mbox{g}\cdot \mbox{L}^{-1}$ \\
      $O^*$ & $0.015$ $\mbox{g}\cdot \mbox{L}^{-1}$ & & & \\
      $k_La$ & $30.0$ $\mbox{g}\cdot \mbox{L}^{-1}$ & & & \\
      $\sigma$ & $1.0$ & & & \\
      $D_G$ & $0.0$ $\mbox{h}^{-1}$ & & & \\
      $D_P$ & $0.5$ $\mbox{h}^{-1}$ & & & \\
      $D_O$ & $10.0$ $\mbox{h}^{-1}$ & & & \\
    \end{tabular}
  \end{center}
\end{table}

Although the model presented in this section provides a highly simplified description of intracellular metabolism and spatial structure in the environment or population, it has captured some fundamental yet non-trivial aspects of cooperative metabolic resource allocation in microbial systems. One simplification is the assumption of well-defined sub-populations from the outset, which avoids a central question on how such spatial structure actually emerges. The general theory does not require this stricter assumption however, and in principle allows an initially near-homogenous collection of cells to self-organise into distinct, complementary  sub-populations in response to subtle differences in environmental conditions or metabolic thresholds \cite{Varahan19}. Further extensions to accommodate metabolic models with more detailed sets of pathways can be guided using the unique recursive nature of the maximum entropy control law in combination with EFM families as described in \cite{Tourigny20}. Moreover, increased spatial complexity is also readily introduced following the procedures outlined in \cite{Harcombe14,Biggs13,Chen16,Zhang18}. In fact, in \cite{Zhang18} a very similar dynamic model to that presented here was considered along with a rather different community-wide objective based on minimising the enthalpy of combustion in the environment. Finally, even in their current form, the dynamic and structural aspects of the model already apply to various lactate-based metabolic cross-feeding mechanisms hypothesised for higher eukaryotic organisms \cite{Belanger11,Sonveaux08,Semenza08,Lyssiotis17} due to the high level of conservation in central carbon metabolism.   

\section{Conclusion}      

Organisms making regulatory decisions subject to uncertainty in future growth conditions must also contend with having very limited knowledge about their peers in spatially-structured environments and communities. In this paper, the maximum entropy control law that defines a dynamic, bet-hedging strategy for metabolic resource allocation has been extended to encompass local cooperative behaviour of individuals working together to achieve a global, population-wide metabolic objective. Analogous to the way in which the individualistic maximum entropy control predicts accumulation of metabolite reserves under growth-limiting conditions, its cooperative extension similarly describes when it would be optimal to allocate greater resources to pathways involved in metabolic cross-feeding over those that yield exclusive benefit to the individual. As for spatially-homogenous systems, resource fractions allocated to metabolic pathways not directly contributing to instantaneous growth (e.g., formation of storage metabolites or cross-feeding products) will tend to increase as environmental conditions become growth-limiting, but in spatially-heterogenous systems individuals may also take into consideration their current growth status relative to other community members when regulating metabolic cross-feeding. In this way, the maximum entropy principle provides a powerful framework for understanding the bet-hedging nature of dynamic metabolic resource allocation in space and time.                        

\section*{Acknowledgements}
This work benefitted from conversations with HC Causton, L Dietrich, and MR Parsek on microbial biology and quorum sensing. DS Tourigny is a Simons Foundation Fellow of the Life Sciences Research Foundation.
      
\appendix

\section{Derivation of the cooperative control law}
\label{sec:appendix}
Collecting dynamic variables into vectors $\mathbf{X}_i = (\mathbf{m}_i,x_i)^T$, the system (\ref{full}) can be represented by the coupled equations $\dot{\mathbf{X}}_i = \mathbf{F}_i(\mathbf{X}, \mathbf{u}^i)$ ($i=1,2,...,\mathcal{N}$) with $\mathbf{X} = (\mathbf{X}_1, \mathbf{X}_2, ..., \mathbf{X}_{\mathcal{N}})^T$. As in \cite{Young07,Tourigny20}, the individual controls $\mathbf{u}^i$ are determined by using the linearisation of (\ref{full}) about the state $\mathbf{X}(t)$ and reference controls $\{ \mathbf{u}^i_0 \}_{i=1,2,...,\mathcal{N}}$ as a good approximation for the system response at time $t+\tau$ with $\tau \in [0,\Delta t]$. The linearised equations are 
\begin{equation*}
\frac{d}{d\tau} \Delta \mathbf{X}_i = \mathbf{A}_i \Delta \mathbf{X} + \mathbf{B}_i \Delta \mathbf{u}^i + \mathbf{F}_i(\mathbf{X}(t),\mathbf{u}^i_0)
\end{equation*} 
where
\begin{equation*}
\mathbf{A}_i = \frac{\partial}{\partial \mathbf{X}} \mathbf{F}_i (\mathbf{X}(t),\mathbf{u}^i)
\end{equation*}
is the $i$th block row of the full Jacobian matrix $\mathbf{A}$ and
\begin{equation*}
\mathbf{B}_i = \frac{\partial}{\partial \mathbf{u}^i} \mathbf{F}_i (\mathbf{X}(t),\mathbf{u}^i),  
\end{equation*}
with $\Delta \mathbf{X} = \mathbf{X}(t+\tau) - \mathbf{X}(t)$ and $\Delta \mathbf{u}^i = \mathbf{u}^i (t+ \tau) - \mathbf{u}^i_0$. When the species at all nodes are identical (i.e., $\mathbf{S}_i = \mathbf{S}$, $K_i = K$, $\mathbf{c}_i = \mathbf{c}$, $r^i_k = r_k$, and $\mathbf{Z}^k_i = \mathbf{Z}^k$ for all $i = 1,2,..., \mathcal{N}$) then each cooperative maximum entropy control $\mathbf{u}^i$ is obtained by maximising the objective functional 
\begin{equation*}
\mathcal{F}_i(\mathbf{u}) = \boldsymbol{\lambda}^T_i \mathbf{B}_i \mathbf{u} + \sigma H(\mathbf{u})
\end{equation*}  
with respect to $\mathbf{u}$, where 
\begin{equation*}
H(\mathbf{u} ) = - \sum_{k=1}^K u_k \log(u_k)
\end{equation*}
is the entropy constraint and $\boldsymbol{\lambda} = (\boldsymbol{\lambda}_1,\boldsymbol{\lambda}_2,...,\boldsymbol{\lambda}_{\mathcal{N}})^T$ is the $\mathcal{N} \times (M+1)$-dimensional vector of Pontryagin co-state variables obtained by solving the boundary value problem
\begin{equation*}
- \frac{d}{d \tau} \boldsymbol{\lambda} = \mathbf{A}^T \boldsymbol{\lambda}, \quad \boldsymbol{\lambda}(t + \Delta t) = \mathbf{q}_i .
\end{equation*}
The complete solution to the co-state equations is 
\begin{equation*}
 \boldsymbol{\lambda} (t + \tau) = \mathbf{e}^{\mathbf{A}^T(\Delta t - \tau)} \mathbf{q}_i, \quad 0 \leq \tau \leq \Delta t ,
\end{equation*}
but the effective return-on-investment (\ref{eroi}) is obtained by substituting for $ \boldsymbol{\lambda}$ with the heuristic $\tau =0$ because, ultimately, only the optimal control input at the current time $t$ is of interest \cite{Young07}. Maximisation of $\mathcal{F}_i$ proceeds as in \cite{Tourigny20} to yield the cooperative maximum entropy control (\ref{control}).

An expression for the greedy effective return-on-investment is obtained by setting $\Delta t = 0$, in which case $\boldsymbol{\lambda}_i = \partial \phi_i (\mathbf{X})/\partial \mathbf{X}_i$. Moreover, $\partial \phi_i/\partial x_i$ is the only non-zero component of this vector because the cooperative objective $\phi_i$ does not depend on concentrations of slow metabolites. Differentiation of the generalised mean $\phi_i = M_p(\mathbf{x})$ with respect to $x_i$ yields 
\begin{equation*}
\frac{\partial \phi_i}{\partial x_i}  = \frac{x^{p-1}_i}{\sum_{j=1}^\mathcal{N} x_j^p}  \cdot \left( \frac{1}{\mathcal{N}} \sum_{j=1}^\mathcal{N} x_j^p \right)^{1/p} =  \frac{x_i^{p-1}}{x_i^p + y^p} \cdot M_p(\mathbf{x}) 
\end{equation*}   
and the $k$th column of $\mathbf{B}_i$ is 
\begin{equation*}
\mathbf{B}_i^k = x_i r_k(\mathbf{m}_i) \begin{pmatrix}  \mathbf{S}  \\ \mathbf{c}^T  \end{pmatrix}  \mathbf{Z}^k
\end{equation*}
so that
\begin{equation*}
\mathcal{R}^{k,i}_0 = \frac{x_i^p}{x_i^p + y^p} \cdot M_p(\mathbf{x})  \cdot r_k(\mathbf{m}_i)  \mathbf{c}^T \mathbf{Z}^k .
\end{equation*} 
This provides the greedy effective return-on-investment (\ref{geroi}) defined using $R^k_0(\mathbf{m}_i)$ in the main text, and the resulting greedy cooperative maximum entropy control is 
\begin{equation*}
u^i_k(t) = \frac{1}{Q_i}\exp \left( \mathcal{R}_0^{k,i} /\sigma  \right) , \quad k=1,2,...,K
\end{equation*}  
with $Q_i = \sum_{k=1}^K \exp ( \mathcal{R}_0^{k,i} /\sigma)$. As mentioned in the main text, alternative controls can be obtained in cases where the objectives $\phi_i$ and/or species are different across nodes in the population network, but are not considered here. To demonstrate the utilitarian objective with $p=1$ recovers the individualistic greedy maximum entropy control law from \cite{Tourigny20}, note that in this case
\begin{equation*}
M_1(\mathbf{x}) = \frac{1}{\mathcal{N}} \sum_{j=1}^\mathcal{N} x_j , \quad x_i^1 + y^1 = x_i + \sum_{j \neq i }^\mathcal{N} x_j =  \sum_{j=1}^\mathcal{N} x_j
\end{equation*}
so that substitution for the denominator in $\mathcal{R}^{k,i}_0 $ yields
\begin{equation*}
\mathcal{R}^{k,i}_0 = \frac{1}{\mathcal{N}} \cdot x_i r_k(\mathbf{m}_i)  \mathbf{c}^T \mathbf{Z}^k ,
\end{equation*}
which is the individualistic effective return-on-investment (Equation 18 in \cite{Tourigny20}) for the $k$th EFM at node $i$, multiplied by a common factor of $1/\mathcal{N}$ that has no effect on the resulting control law.

\end{document}